\documentclass[a4paper,11pt,oneside]{article}

\author{\large \bf G\"oran Einarsson \\ \\
         Royal Institute of Technology, Telecommunication Theory, \\
         Dept. of Signals, Sensors and Systems, 10044 Stockholm, Sweden. \\
         Phone;  +46 8 7906578, Fax: +46 8 7909370, 
         E-mail: einarsson@s3.kth.se}

\title{\Large \sc Probability Analysis of a Quantum Computer} 
\date{January 15, 2003}

\usepackage{psfrag}
\usepackage{amssymb}
\usepackage{amsmath}
\usepackage{graphicx}
\usepackage{longtable}

\begin{document}

\maketitle                
\section*{Abstract}
The quantum computer (QC) algorithm 
by Peter Shor \cite{Shor 1999} for factorization of integers 
is studied.
The quantum nature of a QC  makes its outcome
random. The output probability distribution is investigated and 
the chances of a successful operation is determined.

\section{Introduction}
To determine the prime factors $p_1$ and $p_2$ of an integer 
$n=p_1 \cdot p_2$ by the Shor algorithm \cite{Shor 1999}
a random integer $1<x<n$ is generated and an estimate of the order 
of $x\ ({\rm{mod}\ n})$ i e. the least positive integer $r$ satisfying 
$x^r \equiv 1\ ({\rm{mod}\ n})$ is determined from the  
QC output state.
The QC contains quantum circuitry calculating the function 
$f(a) = x^a ({\rm mod}\ n)$ and performing a Quantum Fourier Transform (QFT).
It uses two registers one of size $q_A = \lceil 2 \log_2 n \rceil$
qubits, which is read out, and one of size $q_B = \lceil \log_2 n\rceil $ 
containing the function $f(a)$. If $n$ is a $m$ bit number $q_A =2m$.
There are $N = 2^{q_A}$ possible output 
states of register $A$ each of which can be represented a binary number $c$. 
From the QC output $c$ the factors  $p_1$ and $p_2$ can be determined, 
under certain conditions, by operations on $c$ by an ordinary digital
computer 

The critical parameter is the integer $x$ since the quantum circuitry
is constructed for a specific value of $x$.
It is shown in the sequel that an $x$ generated randomly 
from a uniform distribution can be expected to work with probability
2/3, independent of computer size.

\section{Quantum Computer operations}
To factorize a number $n$ by the Shor algorithm a random number $x$
is generated and quantum circuitry  
implementing the function $x^a ({\rm mod}\ n)$ is designed. As a first step
a superimposed state is produced
\begin{equation}
\label{eq c1}
 | \Psi_1 \!\!> = 
 \frac{1}{\sqrt{N}} \sum_{a=0}^{N-1}| a \!\!>|x^a ({\rm mod}\ n)\!\!>
\end{equation}
Applying the QFT to the state $| a \!\!>$  yields
\begin{equation}
\label{eq c2}
| a \!\!>\ \Rightarrow\ 
    \frac{1}{\sqrt{N}} \sum_{b=0}^{N-1} e^{2 \pi i a b /N }\,| b \!\!>
\end{equation}
and the state $| \Psi_1 \!\!>$ is transformed into
\begin{equation}
\label{eq c3}
 | \Psi_2 \!\!> = 
 \frac{1}{N} \sum_{a=0}^{N-1}\sum_{b=0}^{N-1}\exp(2\pi i a b/N) 
| b \!\!>|x^a ({\rm mod}\ n)\!\!>
\end{equation}
The state  $| \Psi_2 \!\!>$ is measured in the reference coordinate system.

The probability that a particular state
         $| c,\ x^k (\large {\rm mod}\ n)\!\!>$ is observed is
\begin{equation}
\label{eq c4}
{\rm P}(c,k) = |\frac{1}{N}\sum_{a:x^a = x^k} \exp(2 \pi i a c/N)|^2  
\end{equation}
Since $x$ has order $r$ all $a:0 \leq a \leq N-1$ of the form $a=m r + k$
satisfies $x^a = x^k$. Solving for $m$ gives $m \leq M$ with
$M = \lfloor (N-k-1)/r \rfloor$ where $\lfloor \ \ \rfloor$ denotes
integer part. 
From (\ref{eq c4}) follows
\begin{equation}
\label{eq c5}
{\rm P}(c,k) = |\frac{1}{N}\sum_{e=0}^{M} \exp(2 \pi i (e r + k) c/N)|^2  
\end{equation}
A factor $\exp(2 \pi k c/N)$ of magnitude one can be factored out and   
\begin{equation}
\label{eq c6}
{\rm P}(c,k) = |\frac{1}{N}\sum_{e=0}^{M} \exp(2 \pi i e r c/N)|^2  
\end{equation}
This is a geometric series which can be summed. The result is
\begin{eqnarray}
\label{eq c7}
{\rm P}(c,k) &=& \frac{1}{N^2} \frac{|\exp(2 \pi i (M+1) r c/N)-1|^2}
                  {|\exp(2 \pi i r c/N)-1|^2} \nonumber \\ 
  &=&\frac{1}{N^2} \frac{\sin^2[\pi (M+1) r c/N]}{\sin^2[\pi r c/N]}
\end{eqnarray}

\begin{figure}[tb]
     \includegraphics[width=1.0\textwidth]{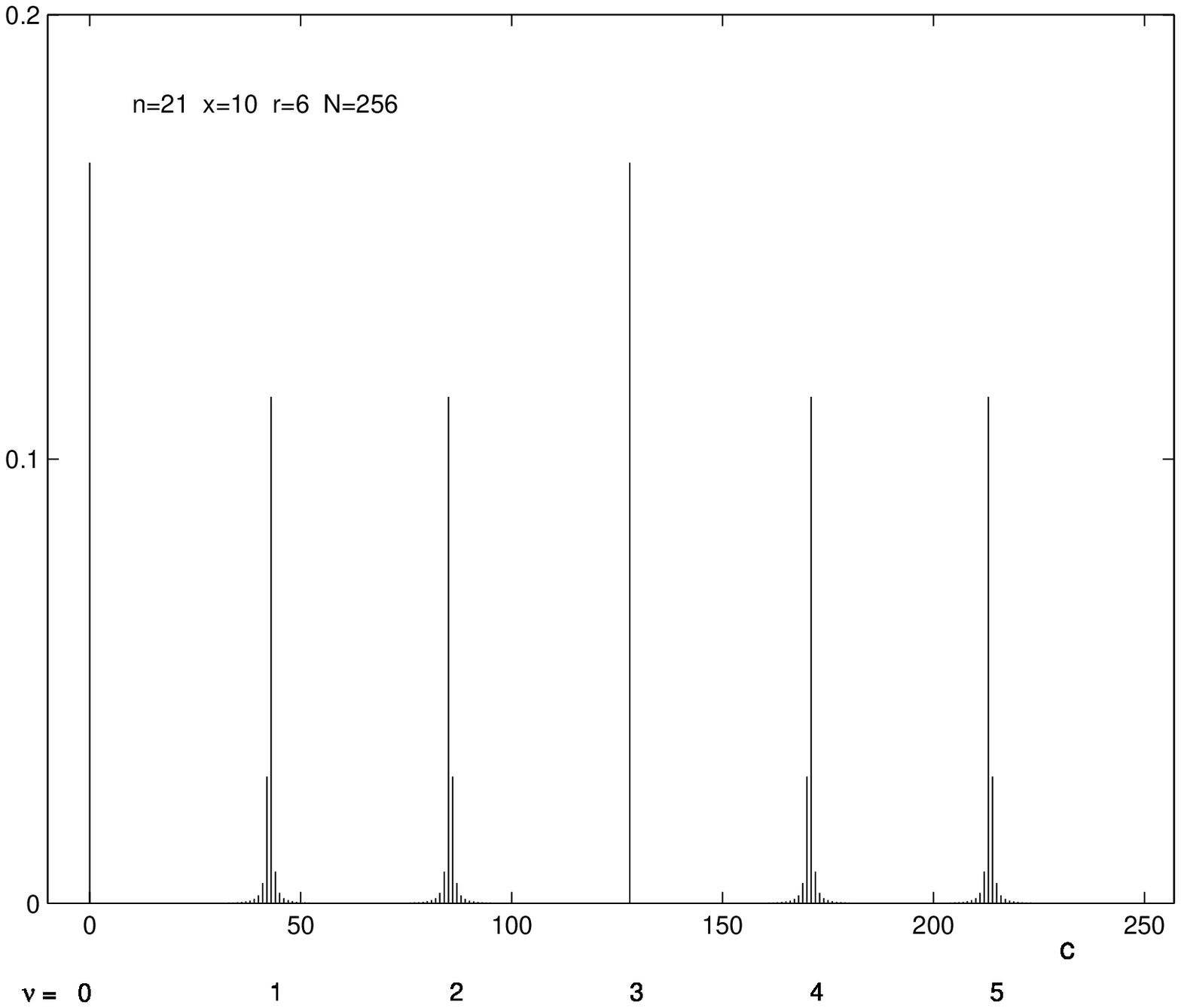}
        \caption{ }
        \label{fig c1}
         Probability distribution $P(c)$ of the Quantum Computer 
         output \\ state number $c$. 
         The integer to factorize $n=21$ and the auxiliary \\ 
         integer  $x=10$. 
         The order of $x\ ({{\rm mod}\ n})$ is $r = 6$. 
         The size of the \\ computer is $q_A = 8$ qubits corresponding to
         $N=256$.
\end{figure}

The probability that the quantum computer ends in state 
$| c \!\!>$ is 
\begin{equation}
\label{eq c12.0}
{\rm P}(c) = \frac{1}{N^2} \sum_{k=0}^{r-1}             
             \frac{\sin^2[\pi (M+1) r c/N]}{\sin^2[\pi r c/N]}
\end{equation}
The parameter $M = \lfloor (N-k-1)/r \rfloor$. Let $k_0$ be the
smallest
integer such that
\begin{equation}
\label{eq c12.1}
\lfloor (N-k_0-1)/r \rfloor = \lfloor (N-r)/r \rfloor = M_0
\nonumber
\end{equation}
Then
\begin{equation}
\label{eq c12}
{\rm P}(c) = \frac{k_0}{N^2} \  
             \frac{\sin^2[\pi (M_0+1) r c/N]}{\sin^2[\pi r c/N]} +
             \frac{r-k_0}{N^2} \ 
             \frac{\sin^2[\pi M_0 r c/N]}{\sin^2[\pi r c/N]} 
\end{equation}
$P(c)$ is illustrated  for $n = 21 = 3 \cdot 7$ and
$x = 10$ in Fig. \ref{fig c1}. The QC register size
$q_A = \lceil \log_2 n\rceil = 9$ making $N=256$

\begin{figure}[tb]
       \includegraphics[width=1.0\textwidth]{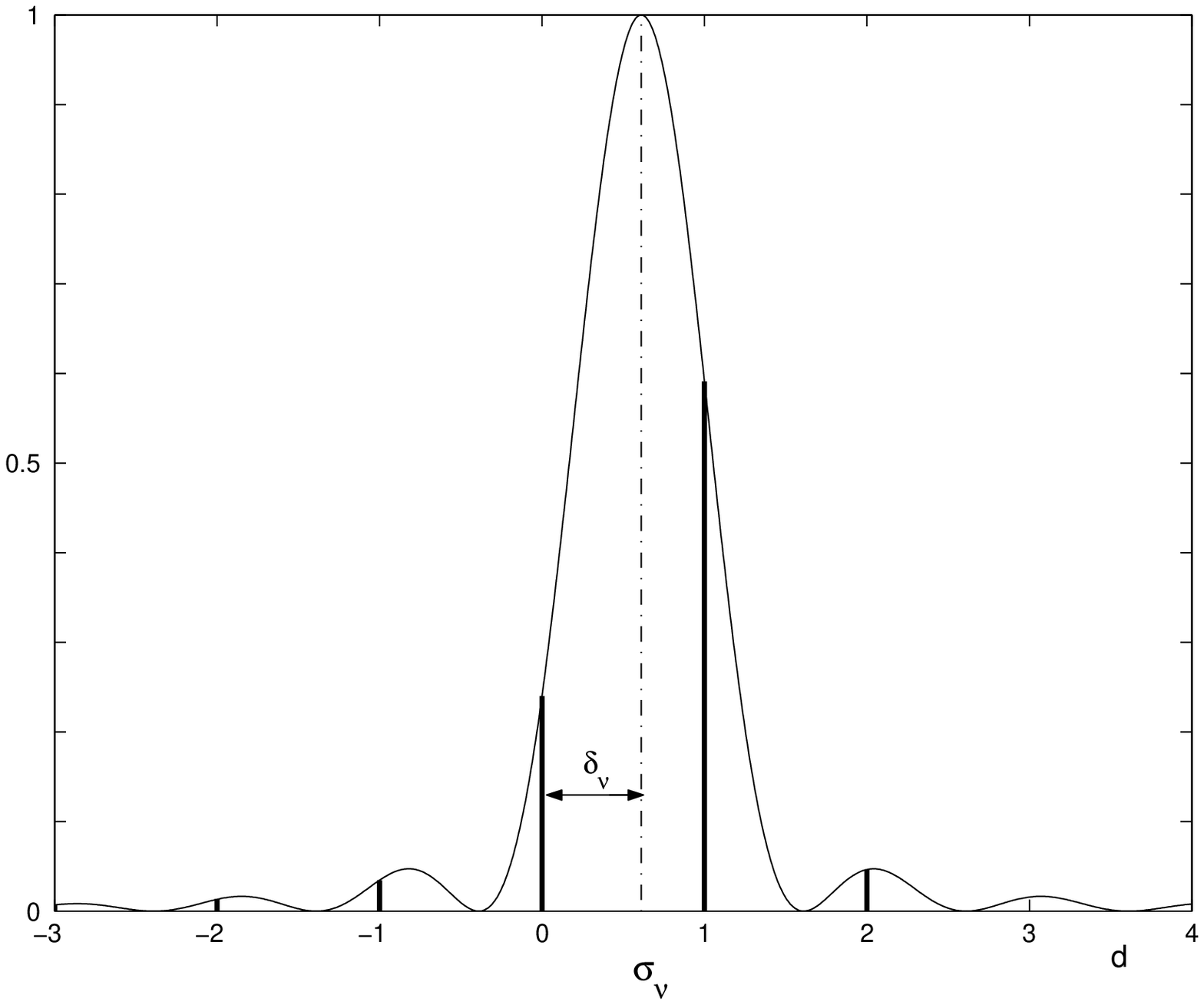}
      \caption{ }
        \label{fig c2}
        The shape of the peaks of the probability distribution
        $P(c)$. \\
\end{figure}

The function (\ref{eq c12}) with $c$ replaced by a real continuous 
variable $\sigma$ is the envelope of the discrete probability distribution.
It is periodic and has peaks at
\begin{equation}
\label{eq c8}
      \sigma_\nu = \frac{\nu}{r} N; \ \ \ \nu = 0, 1, \ldots r-1
\end{equation}
Let $c_\nu$ denote the integer part of $\sigma_\nu$
\begin{equation}
\label{eq c9}
c_\nu = \lfloor \sigma_\nu \rfloor = \sigma_\nu - \delta_\nu; \ \ \ ;
0 \leq \delta_\nu < 1
\end{equation}
The parameters are illustrated in Fig. \ref{fig c2}.

The output $c = c_\nu + d$ from the quantum computer is characterized by
two random variables, the displacement parameter $\delta_\nu$,
and the integer valued parameter $d = c - c_\nu$ representing the 
deviation within the peak.

\section{Post quantum calculations}

The ratio $r/\nu$ is estimated as the largest convergent with nominator 
less than $n$ in a continued fractions expansion of $N/c$.
This procedure generates the fraction $r/\nu$ closest to $N/c$,
see e.g. \cite{Hardy & Wright 1979}, Theorem 181.

The algorithm is based on the relation
\begin{equation}
\label{eq c10.1}
     x^r -1 = (x^{r/2} + 1) (x^{r/2} - 1) \equiv 0 \ ({\rm mod}\ n)
\end{equation}
which shows (means?) that the factors $p_1$ and $p_2$ are possible 
divisors of $(x^{r/2} \pm 1)$.

To obtain a correct value of $r/ \nu$ the 
ratio $c/N$ must not be too far from $\nu /r$
The incorrect value closest the correct $\nu /r$ is $(\nu + 1)/(r + 1)$.
The difference 
\begin{equation}
\label{eq c11}
\Delta = \frac{\nu + 1}{r + 1} - \frac{\nu}{r} = \frac{r - \nu}{r(r + 1)} 
\end{equation}
takes its lowest value $\Delta = 1/r(r + 1)$ for $\nu = r - 1$.
The maximal  value of $r$ is less than $n-1$ and  
$\Delta_{min} > 1/(n-1)n $  
The difference between the correct $\nu/r$ and $c/N$ is
\begin{equation}
\label{eq c11.1}
\Delta_c(d) = \Big\vert \frac{c}{N} - \frac{\nu}{r} \Big\vert = 
              \Big\vert \frac{d - \delta_\nu}{N} \Big\vert  
\end{equation}
For $N \geq n^2$ the distances $\Delta_c(0)$ and $\Delta_c(1)$
are both less than $\Delta_{min}$ and the continued fractions expansion 
always produces the correct order $r$, when $d = 0$ or $d = 1$.

Substitution of $c = \nu N/r + d - \delta_\nu$ into (\ref{eq c12.0})
yields
\begin{equation}
\label{eq c11.2}
{\rm P}(c) = \frac{1}{N^2} \sum_{k=0}^{r-1}             
             \frac{\sin^2[\pi (M+1)\nu + \pi (M+1)r(d-\delta_\nu)/N]}
                  {\sin^2[\pi\nu + \pi r(d-\delta_\nu)/N]}
\end{equation}
The factor $(M+1)r/N = 1-\epsilon$ with $\epsilon < r/N < 1/n$
and since r/N is small 
\begin{equation}
\label{eq c11.3}
{\rm P}(d) = \frac{\sin^2(\pi \delta_\nu)}
                          {{\pi}^2 (d - \delta_\nu)^2}
\end{equation}
is an accurate approximation.
The probability, averaged over
$\delta_\nu$, that  $d = 0$ or $d = 1$ is equal to 0.902.

For more common values of $r$ and $\nu$ the
probability of an incorrect result is much smaller and the continued 
fractions procedure can safely be assumed to produce a  
ratio $\nu/r$ with $r$ is the correct order of $x\ ({\rm mod}\ n)$
and $\nu$ an unknown random number  $\nu = 1,2, \ldots r-1$.
Trying neighboring states, as has been suggested
in the literature, i.e \cite{Shor 1999} p 320, will have no effect.

The relation (\ref{eq c10.1}) requires $r$ to be even and since all even
values of $\nu$ will have at least a factor 2 in common with $r$ the 
probability that $r$ and $\nu$ are relatively prime and the correct 
order $r$ is obtained directly is less than 0.5. 

The algorithm fails when $\nu = 0$, the probability for this
is equal to $1/r$ and can usually be neglected.

\section{Generation of the integer $x$}
The properties of $x$ is of fundamental importance for the function of
the prime factorization algorithm. 
As shown below, for a random $x$ and $n$ the probability
that the factorization procedure fails is 1/3. 

Consider two prime numbers $p_1$ and $ p_2 $ together with an arbitrary 
integer $x$.
Let $r_1$ and $r_2$ denote the order of $x$ mod $p_1$ and mod $p_2$  
respectively.
The order $r$ of  $n = p_1 p_2$ is equal to  ${\rm lcm}(r_1 r_2)$, the least
common multiplier of $r_1$ and $r_2$.
The factorization procedure fails when $r$ is odd i.e when both  
$r_1$ and $r_2$ are odd.
With the assumption that a random $x$ generates random and independent 
values of $r_1$ and $r_2$ and the probability of both being odd is 
$P(A) = 1/4$.

The only other case when the procedure fails is when 
\begin{equation}
\label {eq c15}
                     x^{r/2} \equiv -1 \ ({\rm mod}\ n)
\end{equation}
in which case ${\rm gcd}(x^{r/2}-1) = 1$ and ${\rm gcd}(x^{r/2}+1) = n$. 

The relation (\ref{eq c15}) is satisfied only if both $r_1$ and $r_2$
are even and containing identical factors $2^{k}$ which for random
$r_1$ and $r_2$ occurs with 
probability 
\begin{equation}
\label {eq c16}
       P(B) = \frac{1}{4} \sum_{k=1}^{\infty} \frac{1}{2^{2 k}} 
            = \frac{1}{12}
\end{equation}
making the total probability of failure $P(A) + P(B) = \frac{1}{3}$

Ekert and Jozsa \cite{Ekert & Jozsa 1996} assume $n$ fixed but
arbitrary and showed that 
\begin{displaymath}
Pr[r\ {\rm odd}\ or\ x^{r/2} \equiv -1 \ ({\rm mod}\ n)] \leq 1/2
\end{displaymath}
If $x$ contains any of the prime factors of  $n$ the equation
$x^r \equiv 1\ ({{\rm mod}\ n})$ has no solution and the algorithm fails.
The probability for this is extremely small for any sizable values of $n$.
It is easy to test ${\rm gcd}(x,n)$ before using $x$ and
if there is a common factor no QC calculation is needed. 

\section{Conclusions}
For a quantum computer of size $q_A = 2m$ factorizing an $m$ bit 
number the output from the post quantum calculations 
is the correct order $r$ or $r_{out} = r/ \mu$ with $\mu$ a factor
common to $r$ and $\nu$.
It is easy to check if $x^{r_{out}} \equiv 1\ ({\rm mod}\ n)$ and if not 
the unknown $\mu$ may be found by trial or possibly by a more a
efficient algorithm. From simulation results
\footnote{Quantum Computer simulation programs are available at \\
          http//www.s3.kth.se/$\sim$einarson/}
it has been observed that one of the prime factors can
sometimes be obtained by the gcd algorithm from $r/ \mu$.An
alternative is to rerun the QC with the same $x$to generate a new output $c$.

If the final value of $r$ is either odd or
satisfies (\ref{eq c15}) the algorithm fails and the QC has to be
run with a new $x$, which means that the quantum circuitry needs
to be modified.

\subsection*{Acknowledgment}
Thanks are due to Johan H{\aa}stad for helpful discussions.

\end{document}